# Designing Traceability into Big Data Systems

R McClatchey, A Branson, J Shamdasani & Z Kovacs
Centre for Complex Cooperative Systems
University of the West of England, Bristol BS16 1QY
United Kingdom
Richard.mcclatchey@uwe.ac.uk

The CRISTAL-iSE Consortium
P Emin, M1i, Annecy
P Bornand, Alpha3i Rumilly
France
Patrick.Emin@agilium.com, PBornand@alpha3i.com

*Abstract*—Providing an appropriate level of accessibility and traceability to data or process elements ('Items') in large volumes of data, often Cloud-resident, is an essential requirement in the Big Data era. Enterprise-wide data systems need to be designed from the outset to support usage of such Items across the spectrum of business use rather than from any specific application view. The design philosophy advocated in this paper is to drive the design process using a so-called 'description-driven' approach which enriches models with meta-data and description and focuses the design process on Item re-use, thereby promoting traceability. Details are given of the description-driven design of big data systems at CERN, in health informatics and in business process management. Evidence is presented that the approach leads to design simplicity and consequent ease of management thanks to loose typing and the adoption of a unified approach to Item management and usage.

## I. Introduction

In the age of the Cloud and Big Data, systems must be increasingly flexible, reconfigurable and adaptable to change in order to respond to enterprise demands. As a consequence, designing systems to cater for evolution is becoming critical to their success. To be able to cope with change, systems must have the capability of reuse and the ability to adapt as and when necessary to changes in requirements. Allowing systems to be self-describing is one way to facilitate this and there have been some significant advances recently in systems design which enables us to start to build self-describing systems (which could in time also become self-monitoring and ultimately self-healing) based on the concepts of meta-data, metamodels and ontologies.

Traditionally large data systems have been designed from a set of requirements for system use, as determined from a user community, for a specific business purpose, for example human resource management, inventory control or business information management. These systems evolve over time; new requirements emerge including the need to co-exist with legacy systems and/or to support new activities in an organization. Enterprise system development has helped us take a more holistic view on systems design so that multiple functions in an organization can be supported in a single system. However, over time emerging requirements can subject these systems to frequent change leading to problems with schema evolution in the underlying models and consequently periods of downtime in the operation of the enterprise system.

Related efforts to tackle the problem of coping with design evolution have included, design versioning [1], 'active' object models [2] and schema versioning [3]. However, none of these approaches enables the design of an existing system to be changed *dynamically* and for those changes to be reflected in a new running version of that design. We advocate a design and implementation approach that is holistic in nature, viewing the development of modern object-oriented software from a systems standpoint. It is based on the systematic management of the description of essential systems elements (so-called 'Items') facilitating multiple views of the system under design using pure object oriented techniques.

To address the issues of reuse in designing evolvable systems, this paper proposes a so-called *description-driven approach* to systems design. The exemplar of this approach is our CRISTAL project [4]. CRISTAL is based on description-driven design principles; it uses versions of stored descriptions to define versions of data (or processes) which can be stored in multiple concurrent forms and it is outlined in this paper. We shall show that this approach enables new versions of data structures and processes to be created alongside the old, thereby providing a history of changes to the underlying data models and enabling the capture of *provenance* information. Provenance information includes data on the use of system Items (e.g. data, process or agent Items) and how they have changed over time, by whom and for what purpose thus providing a fine granularity in traceability of the use of Items over the lifecycle of the big data system in question.

The dynamic and geographically distributed nature of Cloud computing makes the capturing and processing of provenance information a major research challenge [5]. To date provenance gathering systems and techniques have mostly been used within scientific research domains such as neuroscience [6] or in bioinformatics [7] but we have also investigated its use with commercial partners for business process management [8]. The usefulness of provenance collection has been discussed at length elsewhere and the interested reader is directed to other works such as [9]. We have developed a concrete application of provenance management in industry which can be harnessed in Big Data system design; it is discussed in this paper.

The structure of the paper is as follows. The next section introduces description-driven concepts and describes the CRISTAL software architecture. In section II we examine the use of CRISTAL for managing a Big Data application in engineering with its use in supporting the construction of the

Compact Muon Solenoid (CMS) experiment at CERN's Large Hadron Collider (LHC). Sections III and IV contrast this with CRISTAL's use in commercial Business Process Management and in supporting clinicians' analyses of MRI images in the search for biomarkers of Alzheimer's disease. In the final section of this paper we evaluate the common design approach underpinning these applications, that of designing flexibility and traceability into these big data applications through the use of description-driven techniques and outline conclusions and future work.

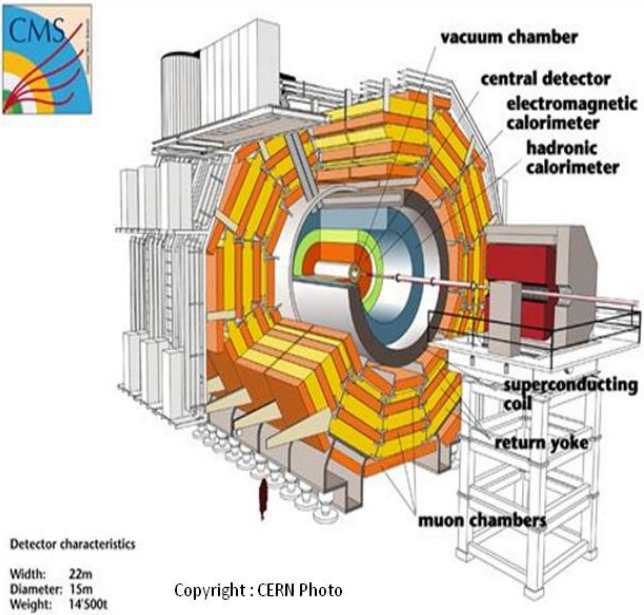

Figure 1. The CMS Detector at the CERN Large Hadron Collider.

## II. DESCRIPTION DRIVEN SYSTEMS AND PROVENANCE

Description-driven systems (DDS) design involves identifying and abstracting, at the outset of the design process, all the crucial elements (such as business objects, processes, lifecycles, goals, agents and outputs) in the system under consideration and creating high-level descriptions of these elements which are stored in a model, dynamically modified and managed separately from their instances. In many ways adhering to a description-driven approach means following very closely the original, and these days often neglected, principles of pure object-oriented design especially those of reuse, abstraction, and loose coupling.

A DDS [10] makes use of so-called meta-objects to store domain-specific system descriptions, which control and manage the life cycles of meta-object instances, or domain objects. In a DDS, descriptions are managed independently to allow the descriptions to be specified and to evolve asynchronously from particular instantiations of those descriptions. Separating descriptions from their instantiations allows new versions of items to coexist with older versions. This separation is essential in handling the complexity issues facing many big data computing applications and allows the realization of interoperability, reusability and system evolution since it gives a clear boundary between the application's basic functionalities from its representations and controls.

The main strength of such a "description driven" approach is that users who develop models of systems need only define them once to create a usable application. The description-driven system then orchestrates the execution of the processes defined in that model (with the consequent capture of provenance information). These descriptions can be modified at runtime and can capture almost any domain; this flexibility has been proven by the development and use of the CRISTAL software in the construction of the CMS ECal [11] at CERN (see figure 1), its application to the Business Process Management (BPM) domain (to model business-based process workflows) and in the manufacturing domain (to control manufacturing processes) and is currently being applied to the HR domain allowing users to modify defined processes.

Scientists at CERN build and operate complex accelerators and detectors whose construction processes are very data-intensive, highly distributed and ultimately require a computer-based system to manage the production, assembly and calibration of components. In constructing detectors like the Compact Muon Solenoid [11] scientists require data management systems that can cope with complexity, with system evolution over time (primarily as a consequence of changing user requirements and extended development timescales) and with system scalability. They also require a very fine granularity of provenance gathering and management over extended timescales. In the case of CMS Electromagnetic Calorimeter (ECal) the construction process took over 10 years with data collected in CRISTAL for eight years up to 2008. CMS has been taking data since at CERN's Large Hadron Collider (LHC).

The ECal construction process was very data-intensive, Grid-resident and highly distributed and its production models changed over time. Detector parts of different model versions had to be handled over the complete construction and usage lifecycle and to coexist with other parts of different model versions. Separating details of model types from the details of parts allowed the model type versions to be specified and managed independently, asynchronously and explicitly from single parts. Moreover, in capturing descriptions separate from their instantiations, system evolution could be catered for while production was underway and provide continuity in the production process and for design changes to be reflected quickly into production, thereby aiding the gathering of historical data. The CRISTAL project was initiated to facilitate the management of the engineering data collected at each stage of production of CMS ECal. CRISTAL is a distributed product data and workflow management system which makes use of an OO-like database for its repository, a multi-layered architecture for its component abstraction and dynamic object modelling for the design of the objects and components of the system [12]. The DDS approach has been followed to handle the complexity of such a data-intensive system and to provide the flexibility to adapt to the changing usage scenarios which are typical of any research production system. Lack of space prohibits detailed discussion of CRISTAL; a full description can be found in [4].

The design of CRISTAL required adaptability over extended timescales for schema evolution, interoperability, deferred commitment and for reusability. In adopting a DDS approach the separation of object instances from object description instances was needed. This abstraction resulted in the delivery of a three layer description-driven architecture (see figure 2). Our CRISTAL approach is similar to the familiar model-driven design concept [13], but differs in that the descriptions and the instances of those descriptions are implemented as objects (Items) and most importantly, they are implemented and maintained using exactly the same internal model. Even though workflow descriptions and instance implementations are different, the manner in which they are stored and are related to each other is the same in CRISTAL.

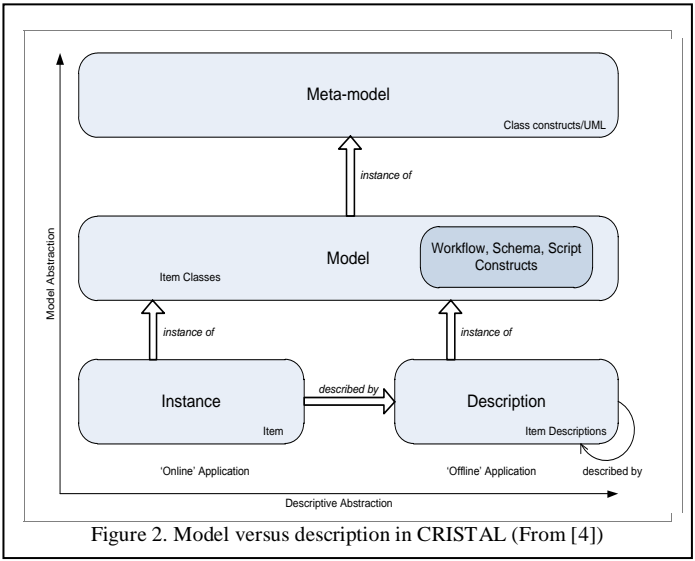

Figure 2. Model versus description in CRISTAL (From [4])

This approach is similar to the distinction between Classes and Objects in the original definition of object oriented principles [14]. We have followed those fundamental principles in CRISTAL to ensure that we can provide the required level of flexibility, maintainability and reusability to facilitate system evolution and the consequent gathering of provenance information at the level of CRISTAL 'Items'.

Within CRISTAL every defined element (or Item) is stored and versioned. This allows users of the system to view older versions of their Items (akin to Objects in Object Orientation) at a later date and either extend a version of an Item or return to a later version of an Item. A full description of the CRISTAL provenance model is out of the scope of this paper however, for clarity the notion of an Item is briefly elaborated upon. CRISTAL is an application server that abstracts all of its business objects into workflow-driven, version-controlled 'Items' which are instantiated from descriptions stored in other Items and are managed on-the-fly for target user communities. Items contain (see figure 3):

• Workflows, which are complete layouts of every action that can be performed on that Item, connected in a directed graph that enforces the execution order of the constituent activities.

• Activities capture the parameters of each atomic execution step, defining what data is to be supplied and by whom. The execution is performed by agents.

• Agents are either human users or mechanical/computational agents (via an API), which then generate events.

• Events detail each change of state of an Activity. Completion events generate data, stored as outcomes. From the generation of an Event provenance information is stored.

• Outcomes are XML documents resulting from each execution (i.e. the data from completion Events), for which viewpoints arise.

• Viewpoints refer to particular versions of an Item's Outcome (e.g. the latest version or, in the case of descriptions, a particular version number).

• Properties are name/value pairs that name and type items. Properties also denormalize collected data for more efficient querying, and

• Collections enable items to be linked to each other.

The basic functionality of CRISTAL is best illustrated with an example: using CRISTAL a user can define product types (such as Newcar spark plug) and products (such as a Newcar spark plug with serial number #123), workflows and activities (test that the plugs work properly, and mount them into the engine). This allows products that are undergoing workflow activities to be traced and, over time, for new product types (e.g. improved Newcar spark plug) to be defined which are then instantiated as products (e.g. updated Newcar spark plug #124) and traced in parallel to pre-existing ones. The application logic is free to allow or deny the inclusion of older product versions in newer ones (e.g. to use up the old stock of spark plugs). Similarly, versions of the workflow activities can co-exist and can be run on these products.

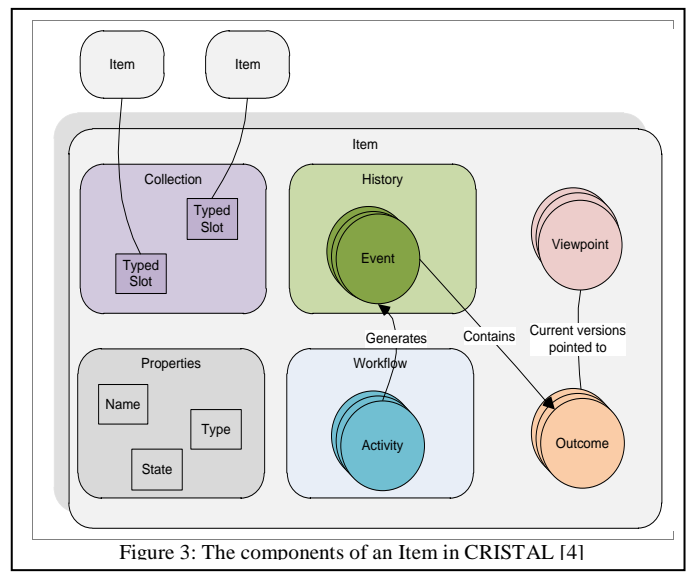

Figure 3: The components of an Item in CRISTAL [4]

In practice some developers find the abstraction concepts of CRISTAL conceptually difficult to understand. This is due to the large amount of terminology involved in the design of CRISTAL as well as the complexity of its concepts. New personnel faced a steep learning curve before they can usefully contribute to the code-base, though this is not a problem for end-users, as complexity may be hidden in intermediate description layers. However, we feel that Items represent a return to the core values of object orientation, at a time when modern languages are becoming increasingly profligate in their implementation of them in the name of expediency, thereby sacrificing many of the benefits of object orientation. Object-orientation encourages the developer to think about the entities involved in the system and the operations required to provide the system's functionality, along with their context in the data model, which together provide the methods of identified data objects, resulting in an object model. In recent years, newer programming languages have tended to focus on object orientation as a means of API specification, increasing the richness of library specification and maximizing code reuse, but do little to encourage good object oriented design amongst developers. Unfortunately, with the increasing popularity of test oriented development methodologies, developers are encouraged to hack away in a deliver-early-and-often way from which a well-thought out object model rarely emerges.

In contrast with CRISTAL the object model must be designed as a set of Items with lifecycles. While other non-Item oriented software components are possible, they cannot store state in the system without interacting with Item activities, and therefore are encapsulated as Agent implementations. These are considered external to the Item model, with a strictly designed outcome specification stating what they must provide to the system to have successfully completed their function. The activities of an Item's lifecycle are roughly analogous to object oriented methods, since they define a single action performed on that Item. However, it is much harder for an Item's lifecycle design to grow out of control with many unused methods since the lifecycle is defined as a workflow; the activity set must always form a valid graph of activities from the creation of the Item to its completion. This clarity of design through implementation constraints is a return to the intentions of the early object oriented languages such as Smalltalk [15], and the initial restrictions of Java, which discouraged the developer from using mechanisms that could result in unstructured, overcomplicated, un-maintainable code, and steer them towards a core object oriented design with the system logic intuitively partitioned and distributed in a manageable way.

During the six years of near-continuous operation, the CRISTAL software collected about a quarter of a Terabyte of scientific data from 450,000 Items needed for the 70,000 elements comprising the CMS ECal detector. During this period there were 22 CRISTAL kernel rebuilds but thanks to the description-driven nature of its design the system only needed to be upgraded seven times, and of those, just one was an update that caused downtime. This was because some data formats originally designed proved not to be as scalable as required; therefore a client update was required to read the new structures. Conventional big data software development separates the specification phase from the construction and implementation phases. However, when the design is evolving as a result of changing user needs, the development process must be reactive and necessarily iterative in nature. The new requirements from the users need to be implemented by the developer in an incremental fashion so that the new results could be assessed and further changes to the design requested, if needed. CRISTAL allows the user to directly verify the business object workflow design, so the normal progression through implementation and testing can be short-circuited. In other words the users can visualize the overall process to be captured in terms of their own recognizable world objects; this greatly simplifies the analysis and (re-)design process. It is relatively easy for professional users to understand the workflow system in CRISTAL, and the nature of XML based data; these both can be detailed by an application maintainer sufficiently accurately in collaboration with the user or may even be drawn by a proficient user directly.

The application logic that needs to be executed during the workflow will have its functionality conveniently broken down along with the activities. It is then simple to import these definitions into the system where it can be immediately tested for feedback to the users. Improvements can thereby be quickly performed online, often by modifying the workflow of one test item, which then serves as a template for the type definitions. Items subject to the improvements can co-exist with items generated earlier and prior to the improvement being made and both are accessed in a consistent, reusable and seamless manner. All this can be done without recompiling a single line of code or restarting the application server, providing significant savings in time enabling users to work in an iterative and reactive manner that suits their research.

In our experience, the process of factoring the lifecycle and dataset of the new item type into activities and outcomes helps to formalize the desired functionality in the user's mind; it becomes more concrete - avoiding much of the vague and often inconclusive discussion that can accompany user requirements capture. Because it evolved from a production workflow specification driven by user requirements, rather than a desire simply to create a 'workflow programming language', CRISTAL's style of workflow correlates more closely to the users' concept of the activities required in the domain item's lifecycle. The degree of granularity can be chosen to ensure that the user feels it provides sufficient control, with the remaining potential subtasks rolled up into a single script. This is one important aspect of the novel approach adopted during CRISTAL development that has proven of benefit to its end-user community. In practice this has been verified over a period of more than 10 years use of CRISTAL at CERN and by its exploitation as the Agilium product [16] across many different application domains in industry. This is discussed in the next section of this paper.

The main lesson learnt from the CRISTAL project in coping with change was to develop a data model that had the capacity to cover multiple types of data (be they products or activities, atomic or composite in nature) and at the same time was intuitively simple. To do this a disciplined and rigorously applied object-oriented approach to data modelling was required: designers needed to think in a way that would ultimately facilitate system flexibility, would enable rapid change and would ease the subsequent burden of maintenance from the outset of the design process. The design approach that was followed in CRISTAL was to concentrate on the essential enterprise objects and descriptions (items, workflows, activities, outcomes, events, viewpoints, properties and collections) that could be needed during the lifetime of the system no matter from which standpoint that data is accessed.

Thus the system was allowed to be open in design and the elegance of its design was not compromised by being viewed from one or several application-led standpoints (such as

Business Process Management (BPM [17]), Enterprise Application Integration (EAI [18]), Workflow Management Systems (WfMS [19]) or whatever. Rather we enabled the traceability of the essential enterprise objects over the lifetime of the system as the primary goal of the system and left the application-specific views to be defined as and when they became required. The ability of description-driven systems to both cope with change and to provide traceability of such changes (i.e. the 'provenance' of the change) we see as one of the main contributions of the CRISTAL approach to building flexible and maintainable big data systems and we believe this makes a significant contribution to how enterprise systems can be implemented. For more detail, consult our previous paper [4] which discusses this in a practical application.

Recently a start-up company called Technoledge [20] has been established to develop applications of CRISTAL that exploit this novelty. Technoledge provide big data provenance management solutions based on a set of customizable software modules that are back-ended by the CRISTAL Kernel for product and process traceability. Technoledge thereby provides both the enterprise repository for capturing business-critical data and the modules for enabling access to and control of that data. Its solutions are generically applicable across business enterprises from scientific and engineering logging applications through manufacturing execution and control to business facing systems for logistics, government, human resources and financial applications. A Technoledge Package is based on the provision of a highly customizable software kernel plus a set of enterprise-agnostic access modules and a customized set of enterprise-specific modules which enable functionality to a particular business enterprise (see figure 4). The Kernel captures the business model for the enterprise to be supported; the modules populate and manage that model and allow access to the critical data held in it so that applications can work with it. The agnostic element comprises reusable modules for the storage, querying, visualization, management, administration and reporting of data supplied by the enterprise-specific modules.

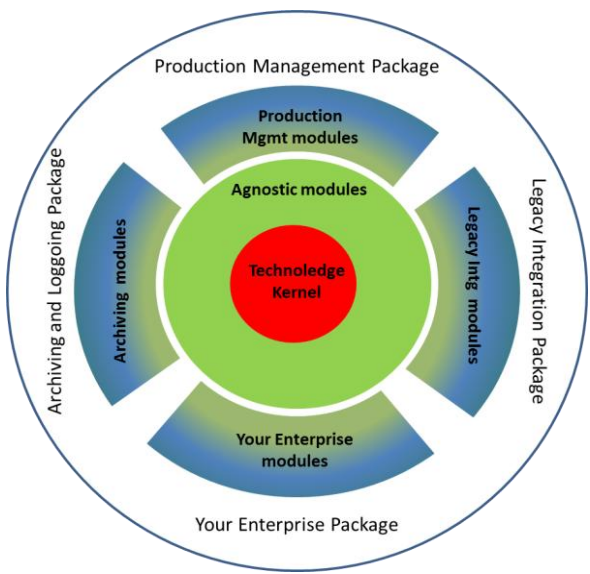

Figure 4. Technoledge use of the CRISTAL Kernel.

Keeping control of changes to big data, or the origins of the changes to the processes involved in capturing this data, is invaluable to any business. The Technoledge suite records this provenance providing a means of capturing and visualizing how clients' enterprise data and processes have changed and evolved. It allows considerable exploration functionality for clients to browse past processes and data descriptions and to instigate change based on those historical records. The overall solution that is offered enables end-users to capture a description of the 'heart' of their enterprise (engineering, finance, retail, manufacturing), to handle process and data logging and via the CRISTAL model to enable system integration with existing systems. The open model and the associated Technoledge modules facilitate business-to-business operation and can be applied across enterprise functions (such as personnel, order management, CRM, and ERP) seamlessly.

III. CRISTAL IN BUSINESS PROCESS MANAGEMENT

Further evidence of the benefits accruing from use of CRISTAL comes from its commercialization as the Agilium product. Since 2004 an early version of the CRISTAL Kernel has been exploited as the Agilium product by the M1i company (based in Annecy, France) for the purpose of supporting BPM and the integration and co-operation of multiple processes especially in business-to-business applications. M1i have taken CRISTAL and added applications for BPM that benefit from the description-driven aspects of CRISTAL, v.i.z. its flexibility and system evolution management. Their product addresses the harmonization of business processes by the use of a CRISTAL database so that multiple potentially heterogeneous processes can be integrated and have their workflows tracked in the database.

Agilium integrates the management of data coming from different sources and unites BPM with Business Activity Management (BAM) [21] and Enterprise Application Integration [18] through the capture and management of their designs in the CRISTAL system. Using the facilities for description and dynamic modification in CRISTAL, Agilium is able to provide modifiable and reconfigurable business workflows. It uses the description-driven nature of the CRISTAL model to act dynamically on process instances already running and can thus intervene in the actual process instances during execution. These processes can be dynamically (re-) configured based on the context of execution without compiling, stopping or starting the process and the user can make modifications directly and graphically of any process parameter. Thus the Agilium system aims to provide the level of flexibility for organizations to be agile in responding to the ongoing changes required by cyber-enterprises, with functionality derived from use of CRISTAL.

The Agilium Server is based on CRISTAL, but with several domain extensions and support for additional protocols added. The user interface (UI) components are the Agilium Web component, the Agilium Supervisor GUI and the Agilium Factory [16]. The Agilium Web is a web application based on J2EE and running within Tomcat as the container. This is where users can browse the currently active jobs and different instances of business processes. The list of jobs available to a user are constrained by their individual roles (for example, administrator). The web UI also allows users to complete manual activities. The supervisor GUI component of Agilium is derived from the original Java Swing CRISTAL GUI, and is used by administrators of the system to be able to design and debug workflows and for general system management. The key component in Agilium is known as the Factory. The Factory is a full Eclipse based application which has a modern UI and allows M1i's users to create and manage their own CRISTAL based workflows. A screenshot of the Agilium Factory is shown in figure 5.

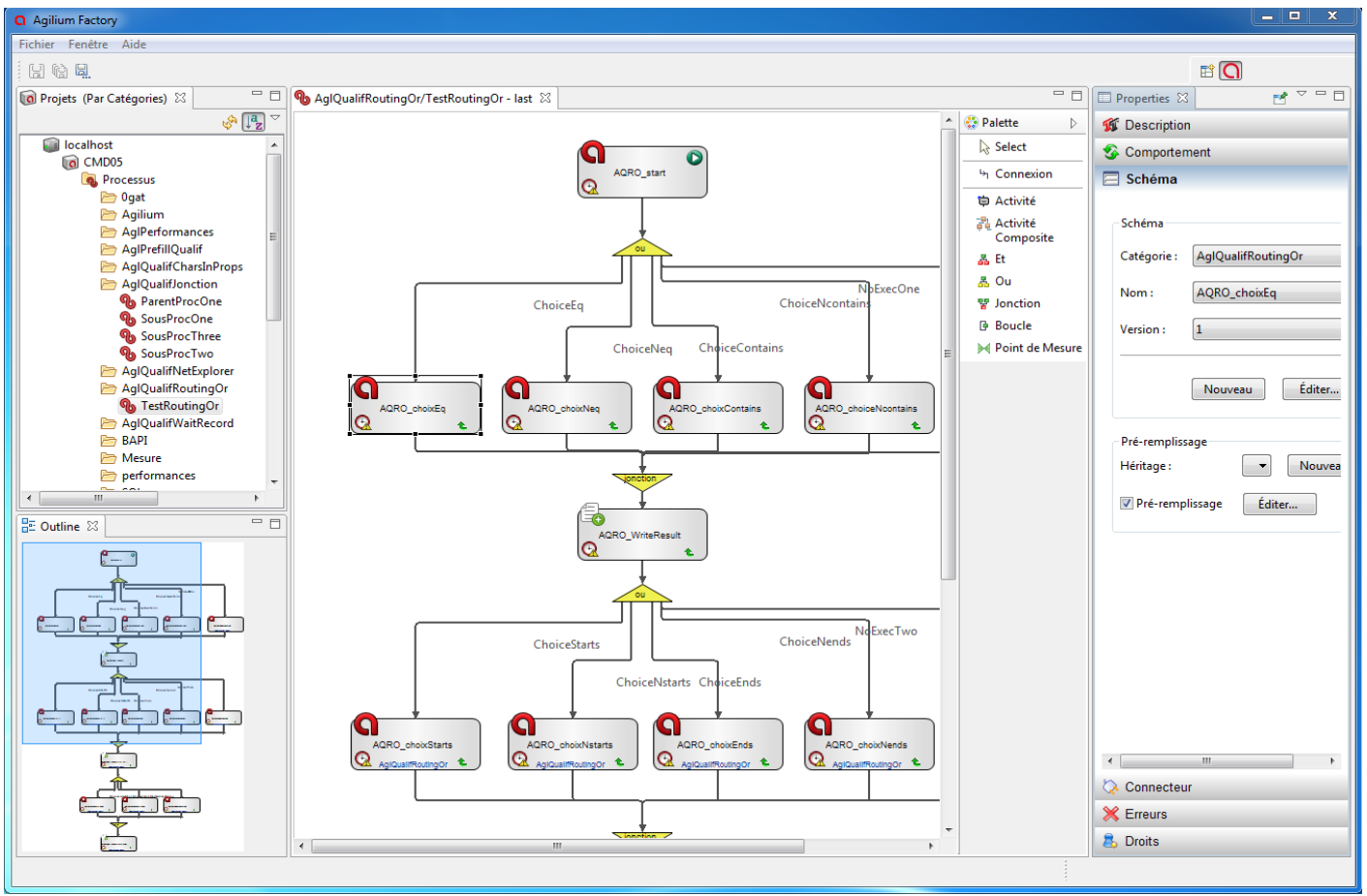

Figure 5: The Agilium Factory Application.

The major benefit to Agilium in the use of CRISTAL is in provenance capture and recording of their Business Process Modelling (BPM) workflow executions. Within the Agilium product, the provenance model is identical to the provenance model of CRISTAL where Events are generated and stored. As stated previously, all models are created at runtime. This means that all BPM workflows developed within Agilium are stored and versioned (and thus their traceability, or provenance, is recorded). This allows users to return at a later date and view previous versions of the BPM models, fix bugs, or to extend their previous BPM workflows in a new design.

One example of where provenance is useful for Agilium is a company which produces solar panels. With this client, the production of each solar panel can take more than a month. They also require different versions of workflows to be stored and accessed on site. Therefore, this client of theirs requires that they be able to look into the past versions of their processes and workflows. This means that they can retrieve the history of all the production steps for each panel, even though the BPM workflow has evolved between the two generations of panels. When an alteration to the fabrication process is required, in the past they have modified their production process to increase the performance level of the solar cells. The workflows corresponding to the production processes are modified to add or remove activities matching an electro-deposition or cleaning step, or to alter their parameters. These modifications are usually done at run time. These changes are saved and stored as newer versions, allowing the panels using the older versions of the workflow to continue unhindered whereas the newer modifications can be applied to newer solar panels in production; this is a key strength of using CRISTAL in Agilium and demonstrates not just the use of provenance but also the flexibility of the system.

The inherent provenance capabilities of CRISTAL mean that the model itself is also versioned, allowing users to look at the production steps for each version of the panels they have created and to see what processes they have in common. This allows them to view and analyze which processes have changed. This aspect is crucial to their business since it allows them to look at the evolution of the production process. The developers at M1i chose CRISTAL as the basis for their system since they felt that its provenance and traceability features were key for them to create a product with a competitive edge in the market. With diversification into Cloud-resident Big Data systems M1i are already realizing the benefits to their advanced BPM solution through the use of mature, proven technology based on the description-driven concepts of CRISTAL.

## IV. ANALYSIS TRACEABILITY WITH CRISTAL

A further application of CRISTAL technologies for big data traceability is that from the neuGRID/N4U EC Framework 7 project studies of medical imaging into Alzheimer's disease. The full details of these studies are beyond the scope of the current paper (details can be found at [22]) but they serve to illustrate the functions of a description-driven system as used for tracing scientific workflows. Scientific workflows are increasingly required to orchestrate research processes in medical analyses, to ensure the reproducibility of analyses and to confirm the correctness of outcomes [23]. In a collaborative research environment, where researchers use each others' results and methods, traceability of the data generated, stored and used must also be maintained. All these forms of knowledge are collectively referred to as provenance information.

In any big data system where there are multiplicities of data-sets and versions of workflows operating upon those data-sets, particularly when the analysis is carried out repetitively and/or in collaborative teams, it is imperative to retain a record of who did what, to which sets of data, on which dates, as well as recording the outcome(s) of the analysis. This provenance information needs to be logged as records of particular users' analyses so that they can be reproduced or amended and repeated as part of a robust research process. All of this information, normally generated through the execution of scientific workflows enables the traceability of the origins of data (and processes) and, perhaps more importantly, their evolution between different stages of their usage. Capturing and managing this provenance data enables users to query analysis information, automatically generate workflows and to detect errors and exceptional behaviour in past analyses.

In the project neuGRID for Users (N4U) we have provided a Virtual Laboratory (VL, see https://neugrid4you.eu) which offers neuroscientists tracked access to a wide range of Cloud-resident big data sets, and services, and support in their study of biomarkers for identifying the onset of Alzheimer's disease. The N4U virtual laboratory, whose architecture is illustrated in Figure 6, is based on services layered on top of the neuGRID infrastructure and a CRISTAL database, described in detail in [24]. The VL was developed for imaging neuroscientists involved in Alzheimer's studies but has been designed to be reusable across other research communities. The VL enables clinical researchers to find clinical data, pipelines, algorithm applications, statistical tools, analysis definitions and detailed interlinked provenance in a user-friendly environment. This has been achieved by basing the N4U virtual laboratory on a so-called integrated Analysis Base (or Data Atlas [24]).

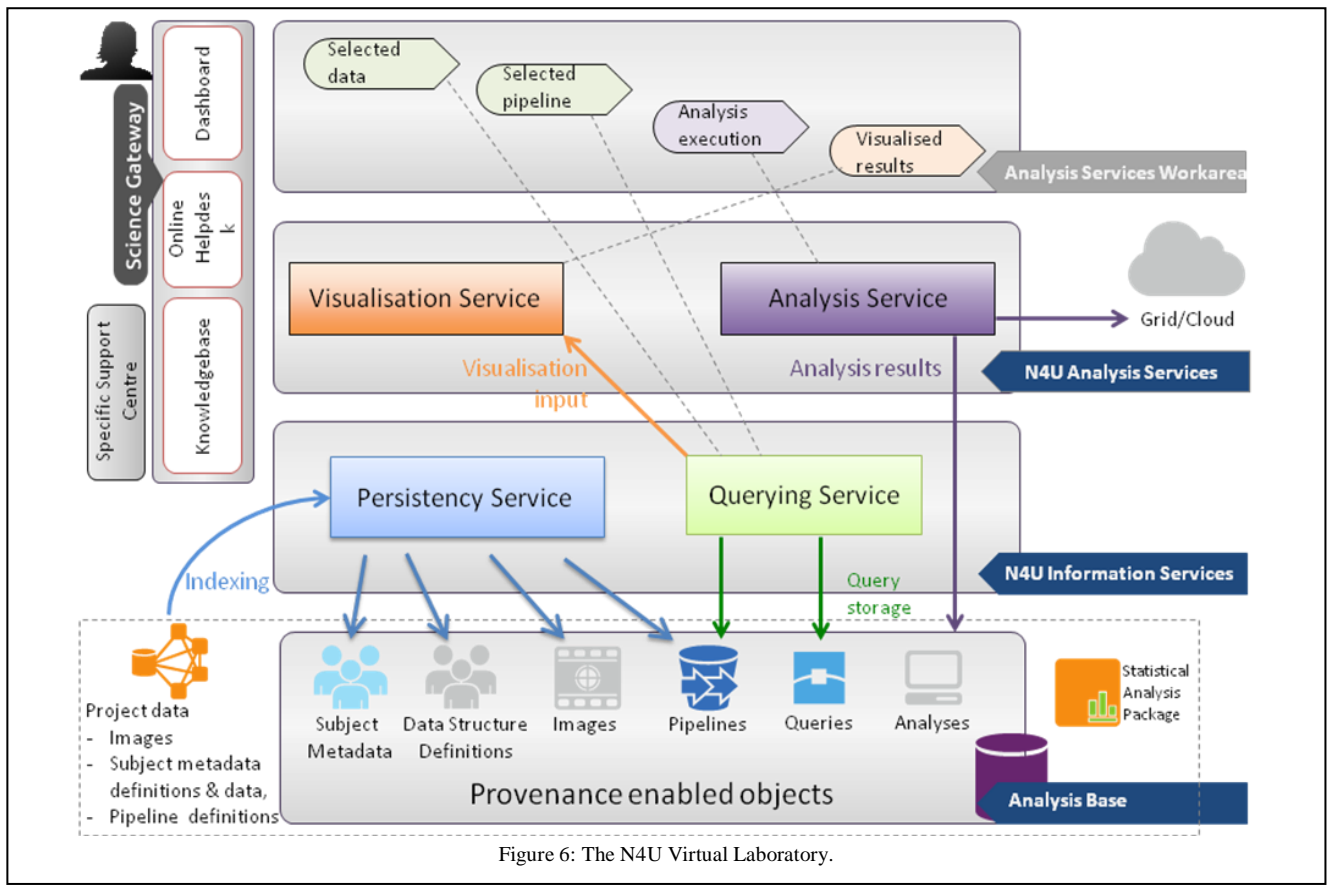

Figure 6: The N4U Virtual Laboratory.

The N4U analysis base addresses practical challenges by offering an integrated data analysis environment to optimally exploit neuroscience workflows, large image datasets and algorithms to conduct scientific analyses. The high-level flow of data and analysis operations between various components of the virtual laboratory and the analysis base are also highlighted in Figure 6. The N4U analysis base enables such analysis by indexing and interlinking the neuroimaging and clinical study datasets stored on the N4U Grid infrastructure, algorithms and scientific workflow definitions along with their associated provenance information.

Once researchers conduct their analyses using this interlinked information, the analysis definitions and resulting data along with the user profiles are also made available in the analysis base for tracking and reusability purposes via a so-called Analysis Service. The N4U virtual laboratory provides the environment for users to conduct their analyses on sets of images and associated clinical data and to have the provenance of those analyses tracked by CRISTAL. In neuGRID/N4U, we have used CRISTAL to provide the provenance needed to support neuroscience analysis and to track individualized analysis definitions and usage patterns, thereby creating a practical knowledge base for neuroscience researchers. The N4U Analysis Service provides access to tracked information (images, pipelines and analysis outcomes) for querying/browsing, visualization, pipeline authoring and execution.

CRISTAL captures provenance data that emerges in the specification and execution of the stages in analysis workflows. The provenance management service also keeps track of the origins of the data products generated in an analysis and their evolution between different stages of research analysis. CRISTAL is a system that records every change made to its objects or Items. Whenever a modification is made to any piece of data, the definition of that piece of data or application logic, the change and the metadata associated with that change (e.g. who made the change, when and for what purpose) are stored alongside that data. This makes CRISTAL applications fully traceable, and this data may be used to assemble detailed provenance information. In N4U, CRISTAL manages data from the Analysis Service, containing the full history of computing task execution; it can also provide this level of traceability for any piece of data in the system, such as the datasets, pipeline definitions and queries.

The Analysis Service provides workflow orchestration for scientists and a platform for them to execute their experiments on the GRID. It allows users to recreate their experiments on the neuGRID/N4U Infrastructure using previously recorded provenance information as well as a set of visualization tools allowing users to view their results and perform statistical analyses. In essence the Analysis Service enables:

• The browsing of past analyses and their results;

• The creation of new analyses by pairing datasets with algorithms and pipelines found in the Analysis Base;

• The execution of analyses by creating jobs to be passed to the Pipeline Service, then logging the returned results in the analyses objects;

• Re-running of past analyses with different parameters or altered datasets and

• The sharing of analyses between researchers.

The detailed operation of the Analysis Service is best understood with a practical example. Consider the case where a clinician wishes to conduct a new analysis. Her first step would be to compile a selection of data from the datasets which are available to her. To do this she would log into the Analysis Service Area and interact with the Querying Service through its user interface to find data that possesses the particular properties she is looking for ((see figure 6). She submits her constraints, which are passed as a query to the Querying Service. The Querying Service then queries the Analysis Base which would return a list of dataset properties and locations which meet her constraints. The Querying Service interface would then display this list to the clinician to approve.

Once the user is satisfied with her dataset selection she combines it with a pipeline specification to create her analysis. To do this she would need to use the Analysis Service Interface to search CRISTAL for existing algorithms that she can use to create a new pipeline or to select a pre-defined pipeline. An analysis is an instantiation of a pipeline in the context of a dataset and a pipeline. Command line utilities will be provided to aid in the creation of a pipeline by connecting different algorithms together as steps. The completed pipeline will have a dataset associated with it. Once this pipeline is ready it will be run on each element of the dataset by CRISTAL.

The pipeline will be sent to CRISTAL which will orchestrate the input pipeline (see figure 6) using a Job Broker and the N4U Pipeline Service. Currently the Pipeline Service is not able to perform workflow orchestration. Therefore a single activity from the input workflow will be sent to the Pipeline Service as a single job using the pipeline API. Once the job has completed, the result will be returned to CRISTAL. Here CRISTAL will extract and store provenance information for this job. This information will contain traceability factors such as the time taken for execution, and whether the job completed successfully. It will store this information internally in its own data model. It will also post this information to the Analysis Base so that this crucial provenance information is accessible by the Querying Service. This loop of sending jobs and receiving the result will continue until the workflow is complete. Once this workflow has completed CRISTAL will once more generate provenance information and store this provenance for the entire workflow in its own internal data store and the Analysis Base. The final result of the completed workflow/pipeline will be presented to the user for evaluation. A link to the completed result in the form of a LFN (a Cloud location) and will be stored in the Analysis Base.

The clinician now has a permanently logged record (provenance data) of her analysis including the datasets and (versions of) algorithms she has invoked, the data captured during the execution of here analysis and the final outcome and data returned by her analyses. These provenance elements may also have associated annotation that she has added to provide further knowledge of her analysis that she or others could consult at a later time to re-run, refine or verify the analysis that has been carried out.

## V. CONCLUSIONS AND FUTURE WORK

The examples described above of the usage of CRISTAL across the spectrum of information systems from High Energy Physics through Medical Imaging to Business Process Management demonstrate the flexibility of the description-driven design approach to systems implementation. It also shows the importance of provenance data capture, management and its use in the traceability of data and in large data volume applications. These techniques are generally applicable to any big data development for Cloud-resident data.

Following a description-driven approach designers concentrate on the important building blocks at the heart of their systems - the mission-critical enterprise objects whose lifecycles (creation, versioning, usage, evolution and termination) require tracing for the business to function properly. These 'Items' may be any of data elements or sets, images, video, composite or atomic activities, agents, roles or people and these are specified at the outset of design in an instance of CRISTAL supporting a flexible, described and extensible data model. As described earlier these Items have workflows which are complete layouts of every action that can be performed on that Item, connected in a directed graph that enforces the execution order of the constituent activities etc. (see section 2).

Items are given descriptions and meta-data is associated with both the Items and their descriptions and the designer considers questions such as when are the Items created and by whom and for what purpose? Who can change them over time and why? What data does each Item generate as outcomes when activities are run against those Items? How can the Item be viewed and for what purpose? Does the Item persist and in which versions can it be used concurrently? Note that here we are referring to the Item and its use across the business model rather than seeing the business model as supporting a specific application. Thus the very same 'Person' Item may be viewed via a Personnel system, a Project Management system, a Payroll system, a Training Management system, a Resource Scheduling system or whatever functions that are critical to the operation of that enterprise.

Following the best principles of pure object-oriented design, especially the ideas underpinning the familiar Model-View-Controller paradigm of Smalltalk [15] and the tried and tested principles of re-use, late binding, polymorphism, deferred commitment and inheritance, description-driven design enables flexibility and traceability of design decisions to be built into the big data models that it can support (as demonstrated by the examples given in this paper). Although the CRISTAL approach seems somewhat abstract in terms of its handling of data instances (and their meta-data), models and metamodels, its implementation turns out to be elegant in its simplicity. Items and their descriptions are defined at whatever level is suitable for system management - data, model or meta-model levels (see figure 2) – but, crucially, are treated in the same way with the same code throughout their lifecycles bringing a level of design consistency, coherence and uniformity.

The studies described in this paper have shown that describing a proposed big data system explicitly and openly from the outset of the project enables the developer to change aspects of it responsively as users' requirements evolve. This enables seamless transition from version to version with (virtually) uninterrupted system availability and facilitates full traceability throughout the system lifecycle. Indeed, the description-driven design approach takes object-oriented design this one step further and provides reuse of meta-data, design patterns and maintenance of items and activities (and their descriptions). Practically this results in a higher level of control over design evolution and simpler implementation of system improvements and easier maintenance cycles.

In practice we have found that many system elements have gained in conceptual simplicity and consequent ease of management thanks to loose typing and the adoption of a unified approach to their online manipulation: activities/scripts and their methods; member types and instances; properties and primitives; items and collections; and outcome schemas and views. One logical consequence of providing such a unified design and simplicity of management is that the CRISTAL software can be used for a wide spectrum of application domains.

Future work is being carried out to model domain semantics e.g. the specifics of a particular application domain such as healthcare, public sector, finance, and aerospace. This will essentially transform CRISTAL into a self-describing model execution engine, making it possible to build applications directly on top of the design, largely without code generation. The design will be the framework for all of the application logic – without the risks of misalignment and subsequent loss that code generation can bring – and for CRISTAL to be configured as needed to support the big data application logic whatever it may be. What this means is that the CRISTAL kernel will be able to capture information about the application area in which a particular instance is being used. This will allow usage patterns to be described and captured, roles and agents to be defined on a per-application basis, and rules and outcomes specific to particular user domains to be managed. This will enable multiple instances of CRISTAL to discover the semantics required to inter-operate and to exchange data.

Research into the further extension and uses of CRISTAL continues. There are plans to enrich its kernel (the data model) to model not only data and processes (products and activities as items) but also to model agents and users of the system (whether human or computational). It is planned to investigate how the semantics of CRISTAL items and agents could be captured in terms of ontologies and thus mapped onto or merged with existing ontologies for the benefit of new domain models. The emerging technology of big data analytics and cloud computing and its application in complex domains, such as medicine and healthcare, provide further interesting challenges. To support this in Q4 2014 a version of CRISTAL called CRISTAL-ISE was released to the public as Open Source under the LGPL V3.0 licensing scheme (see www. http://cristal-ise.github.io/).

In the long run we intend to research and develop a so-called Provenance Analysis module for CRISTAL. This will enable applications built with CRISTAL to learn from their past executions and improve and optimize new studies and processes based on the previous experiences and results. Using machine learning approaches, models will be formulated that can derive the best possible optimisation strategies by learning from the past execution of experiments and processes. This will have particular application in manufacturing execution and big data analysis suites. These models will evolve over time and will facilitate decision support in designing, building and running the future processes and workflows in a domain. A provenance analysis mechanism will thus be built on top of the data that has been captured in CRISTAL. It will employ approaches to learn from the data that has been produced, find common patterns and models, classify and reason from the information accumulated and present it to the system in an intuitive way. This information will be delivered to users while they work on new processes or workflows and will be an important source for their future decision-making and design decision traceability and to support new applications built for the (post-) big data era.

ACKNOWLEDGEMENTS

The authors wish to highlight the support of their home institutes across all of the projects that led to this paper and acknowledge funding from the European Union Seventh Framework Programme (FP7/2007-2013) under grant agreement n. 211714 ("neuGRID") and n. 283562 ("neuGRID for users"). Particular thanks is given to Dr Jean-Marie le Goff of CERN, a co-inventor of the CRISTAL system, to Florida Estrella from UWE and CERN and to all past colleagues who have worked on CRISTAL over the period 1996 to the present.